\documentclass[
superscriptaddress,
 amsmath,amssymb,
 aps,
pra,
twocolumn
]{revtex4-1}

\usepackage{graphicx}
\usepackage{dcolumn}
\usepackage{bm}
\usepackage[unicode=true, breaklinks=false, pdfborder={0 0 1}, backref=false, colorlinks=true, linkcolor=blue, urlcolor=blue, citecolor=blue]{hyperref}

\usepackage{bbm}
\newcommand{\id}{\mathbbm{1}}

\newcommand{\tr}{\operatorname{tr}}
\usepackage{braket}
\usepackage{mhchem}

\begin{document}

\title{Spin-chain model of a many-body quantum battery}

\author{Thao P. Le}
\affiliation{School of Physics and Astronomy, Monash University, Victoria 3800,
Australia}
\affiliation{Department of Physics and Astronomy, University College London,
United Kingdom}

\author{Jesper Levinsen}

\affiliation{School of Physics and Astronomy, Monash University, Victoria 3800,
Australia}
\affiliation{ARC Centre of Excellence in Future Low-Energy Electronics Technologies, Monash University, Victoria 3800 Australia}

\author{Kavan Modi}

\affiliation{School of Physics and Astronomy, Monash University, Victoria 3800,
Australia}

\author{Meera Parish}

\affiliation{School of Physics and Astronomy, Monash University, Victoria 3800,
Australia}
\affiliation{ARC Centre of Excellence in Future Low-Energy Electronics Technologies, Monash University, Victoria 3800 Australia}

\author{Felix A. Pollock}
\affiliation{School of Physics and Astronomy, Monash University, Victoria 3800,
Australia}

\date{\today}

\begin{abstract}
Recently, it has been shown that energy can be deposited on a collection of quantum systems at a rate that scales super-extensively. Some of these schemes for `quantum batteries' rely on the use of global many-body interactions that take the batteries through a correlated short cut in state space. Here, we extend the notion of a quantum battery from a collection of \emph{a priori} isolated systems to a many-body quantum system with intrinsic interactions. Specifically, we consider a one-dimensional spin chain with physically realistic two-body interactions. We find that the spin-spin interactions can yield an advantage in charging power over the non-interacting case, and we demonstrate that this advantage can grow super-extensively when the interactions are long ranged. However, we show that, unlike in previous work, this advantage is a mean-field interaction effect that does not involve correlations and that relies on the interactions being intrinsic to the battery.
\end{abstract}

\pacs{Valid PACS appear here}

\maketitle

\section{Introduction}

The recent push towards the development of quantum technologies can be viewed through the lenses of two driving forces. The first is the increasing miniaturisation of technology, especially electronics, which will soon require us to account for the nontrivial effects that quantum mechanics will have on minuscule components. Traditional thermodynamics cannot describe quantum-scale devices, and a new understanding of concepts such as work, heat, and entropy is required. This has led to the field of quantum thermodynamics which explores new understandings of those quantities and also involves the study of quantum machines such as heat engines and refrigerators~\cite{Millen_2016, Goold_2016, Vinjanampathy_2016, Bender_2000, Scully_2003, Brunner_2014, Uzdin_2015}. A second driving force of quantum technologies is the potential for \emph{advantages} due to  quantum effects in certain applications, such as in quantum sensing, cryptography, and computation. One scenario which features both of these aspects of quantum technologies is that of a possible quantum enhancement in thermodynamic tasks, such as the charging of batteries~\cite{Alicki_2013, Hovhannisyan_2013, Binder_2015, Campaioli_2016, Ferraro2017, Friis2017, Jaramillo_2016}.

Conventional chemical batteries and electrochemical capacitors may be intrinsically composed of quantum components, but their operation is essentially classical in nature. ``Quantum batteries''\textemdash a term first used by Alicki and Fannes~\cite{Alicki_2013}\textemdash seek to use nonclassical effects such as quantum coherence or quantum entanglement to impart an advantage compared with classical batteries. Typically, quantum batteries have been modeled as a collection of $N$ independent and identical subsystems, to which a temporary charging field is applied in order to extract or deposit work. In particular, Alicki and Fannes found that global entangling operations could extract more work from a quantum battery than local operations~\cite{Alicki_2013}. This was further nuanced by Hovhannisyan \emph{et al.}~\cite{Hovhannisyan_2013}, who found that a series of $N$ global entangling operations can extract the maximum work without creating any entanglement in the quantum battery. This scenario corresponds to taking a time-consuming ``indirect path'' such that the quantum battery only traverses the space of separable states. By contrast, the ``direct path'' taken under the action of a global entangling operation \emph{does} generate entanglement during operation. This led to the conjecture that the rate of work extraction\textemdash that is, the \emph{power}\textemdash is linked to quantum entanglement~\cite{Hovhannisyan_2013}. This was supported by Binder \emph{et al.}~\cite{Binder_2015} who showed that $N$ interacting quantum batteries traversing through entangled subspaces can charge $N$ times faster than the same number of non-interacting batteries confined to uncorrelated subspaces (under the restriction that the initial and final states are completely uncorrelated in both cases such that the comparison is meaningful).

It would therefore appear plausible that quantum entanglement can enhance the charging of a quantum battery. On the other hand, Campaioli \emph{et al.}~\cite{Campaioli_2016} have shown that a quantum battery with $N$ highly mixed qubits can jointly charge $N$ times faster than they would charge independently. Yet, the joint state of $N$ qubits can be chosen to be so highly mixed that it is confined to the separable ball, \emph{i.e.}, while the joint state does become correlated, there is no entanglement at any point in the charging procedure. If entanglement is not the resource for quantum speed-up, then what is?

One answer to this question may lie in the structure of the interaction Hamiltonian. All of the analyses described above consider rather optimistic scenarios that involve $N$-body global interactions between all the subsystems in the quantum battery. As such, Ref.~\cite{Campaioli_2016} additionally considered the case where at most $k$ subsystems can interact with each other (interaction order $k$) and each subsystem appears in at most $m$ interaction terms (participation number $m$). With this constraint, it was found that the quantum enhancement to the charging power is at most a constant factor of ${\cal O}\left(m k^{2}\right)$. Thus, the extensive quantum enhancement attained by Binder \emph{et al.}~\cite{Binder_2015} is due to the interaction order, \emph{i.e.}, $k = N$, while $m = 1$. However, in the recent theoretical work by Ferraro \emph{et al.}~\cite{Ferraro2017}, an  enhancement in a solid-state battery was achieved by the coupling together of all $N$ two-level systems in a cavity, \emph{i.e.}, the participation number $m = N-1$, while $k = 2$.

\begin{figure}
\begin{centering}
\includegraphics[width=.9\columnwidth]{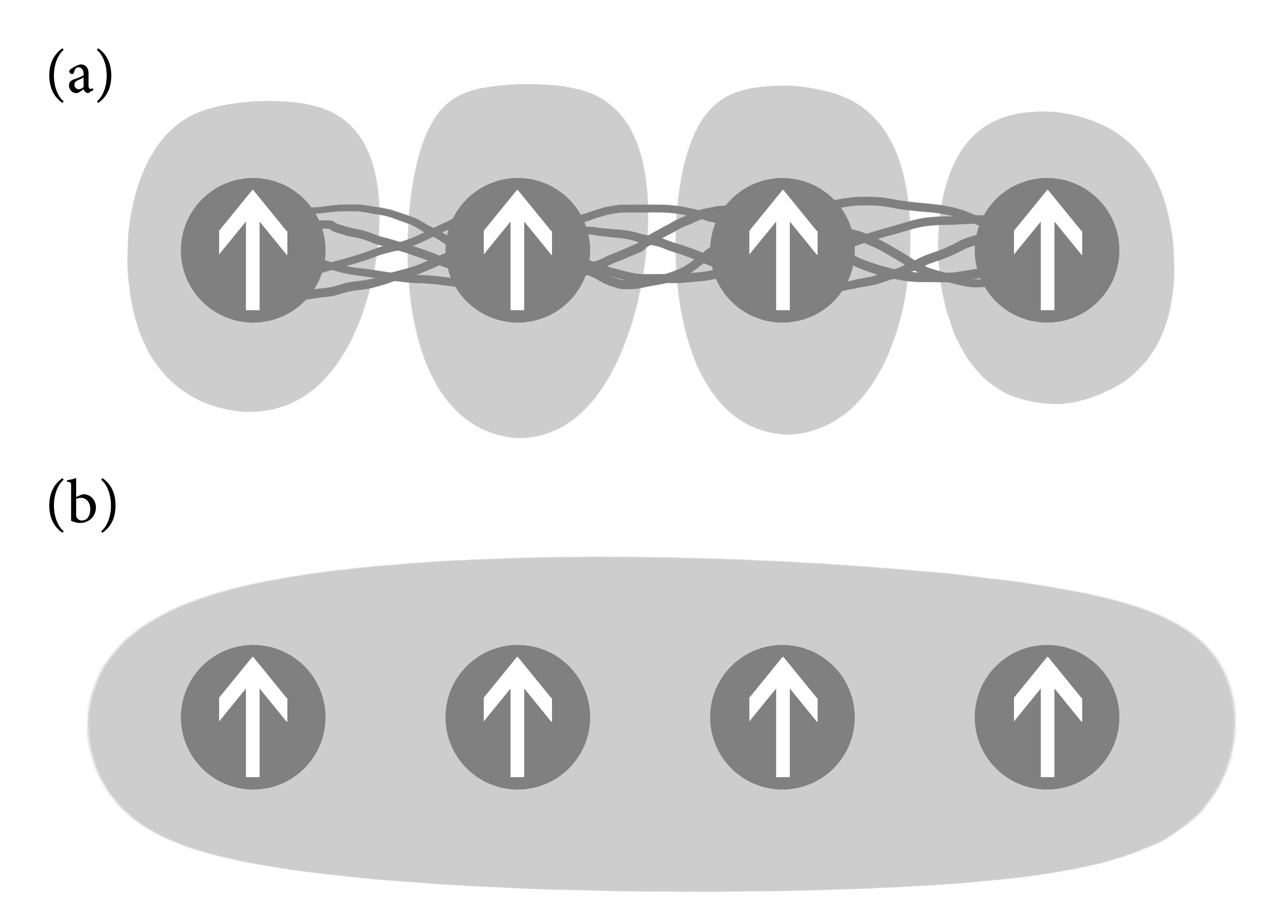}
\par\end{centering}
\caption{(a) A many-body spin-chain battery with internal interactions (represented by bundles of lines), which is charged by \emph{local} charging fields (light shaded regions), \emph{i.e.}, charging fields that in themselves do not couple the spins together. (b) In contrast, previous literature considered independent subsystems charged by a entangling field (large shaded region) that can couple the system together temporarily during the charging process.} \label{fig:spinchainbattery}
\end{figure}

In this paper, we extend the concept of a quantum battery to a many-body quantum system. As opposed to previous works~\cite{Alicki_2013, Hovhannisyan_2013, Binder_2015, Campaioli_2016, Ferraro2017}, where $N$ batteries are jointly charged via global operations, we explore the possibility of locally charging a many-body battery, which can become entangled due to the intrinsic two-body interactions between the system's constituents. Indeed, charging operations which create global entanglement are necessarily emergent operations, while an actual physical system is expected to have a low interaction order. To investigate the possibility of a collective charging speedup in a physically realizable system, we consider a one-dimensional Heisenberg spin chain, which is a fundamental model in condensed matter physics featuring both interactions and the possibility of entanglement. This many-body system has finite interaction order $k=2$ and may have long-range interactions (hence large participation number). Such a deceptively simple Hamiltonian is known to generate arbitrarily complex global entangling operations~\cite{RevModPhys.80.517} and forms a promising basis for certain quantum technologies such as quantum communication~\cite{PhysRevLett.91.207901} and quantum computation~\cite{PhysRevLett.90.247901}. The difference between our spin-chain battery and those considered previously is depicted in Fig.~\ref{fig:spinchainbattery}. Note, though, that the upper bound on charging power derived in Ref.~\cite{Campaioli_2016} for $N$ batteries still holds in our case due to a symmetry between the intrinsic and charging Hamiltonians in the two scenarios~\footnote{Specifically, the upper bound depends on the norm of the commutator of the intrinsic and charging Hamiltonians. Here we have a local charging Hamiltonian and interacting intrinsic Hamiltonian; in Ref.~\cite{Campaioli_2016}, the situation is reversed, but the commutator norm, and hence the bound, does not depend on which is which.}. 

In general, we find that both the interaction range \emph{and} the symmetry of the spin-spin coupling play a role in the work and power derived from the charging process. For isotropic spin-spin interactions (\emph{i.e.}, the coupling is independent of direction) and identical local charging on every spin, we show that the many-body interactions have no effect. However, when the coupling is anisotropic, we find that interactions can provide a boost to the charging power, and the range of the interaction is a major determiner in whether the enhancement is a constant factor, logarithmic, or polynomial in $N$ (as in Ref.~\cite{Binder_2015}). By comparing the quantum evolution to a correlation-less mean-field evolution, we demonstrate that the power is determined by the energetics and interactions of the many-body system. This implies that the correlations that develop during the quantum evolution of our many-body battery are not necessary for enhancing the charging power.

We furthermore demonstrate how an effective Hamiltonian with $N$-body terms emerges naturally as a result of the intrinsic two-body interactions. We show this explicitly for the few-spin problem, where we take advantage of a large spectral gap that develops between a low-energy manifold and higher energy states in the strongly interacting spin chain. In this case, the work deposited onto the battery decomposes into fast and slow oscillations as a function of time, where the slow time scale results from the emergent $N$-body interactions. However, we find that the associated power becomes negligible, and thus this situation does not lead to a quantum advantage.
This raises the question of whether it is ever advantageous to take a direct path through globally entangled subspaces in a physical battery, which is naturally restricted to few-body interactions.

This paper is structured as follows. First, in Section~\ref{sec:battery_model}, we describe our spin-chain battery model and local charging scheme. In Section~\ref{sec:local_charging_results}, we investigate the role of symmetries and determine how effective entangling many-body interactions can be produced in the strong-coupling regime. We also investigate the perturbative limit of weakly interacting spin chains, and in general  how the work deposited and the maximum power depend on the range and isotropy of interactions in the battery. In Section~\ref{sec:classical_chains}, we consider the approximate mean-field evolution of the spin chain, where we neglect correlations, and find that the charging power is comparable to that of the exact quantum evolution. We conclude in  Section~\ref{sec:Conclusions}.

\section{Spin-chain quantum battery}
\label{sec:battery_model}

We consider our quantum battery to be a one-dimensional XXZ Heisenberg spin chain comprised of $N$ spins on a lattice. The spin chain has a precedent of being used in other explorations of quantum devices, communication and computation (\emph{e.g.}~\cite{Galve_2009,Zueco_2009}). Furthermore, spin chains are realised in numerous crystals, such as $\ce{CuCl2 \cdot\text{2} N (C5 D5)}$~\cite{Heilmann_1978}, $\ce{CuGeO_{3}}$~\cite{Hase_1993}, and $\ce{KCuF_{3}}$~\cite{Maillet_2007}, where chains of $\ce{Cu2+}$ along one crystal axis can act as spin chains. Alternatively, spin chains can be engineered using ultracold atoms~\cite{Simon2011,Greif_2013, Murmann2015} or trapped ions~\cite{Smith_2016}.

In the absence of charging operations, we assume that the system has the static Hamiltonian 
\begin{align}
H_{0}  &=H_{B}+H_{g}.    
\end{align}
Here, $H_{B}$ defines an external magnetic field, which acts to break the degeneracy between spins $\ket{\uparrow}$ and $\ket{\downarrow}$: 
\begin{gather}
 H_{B}  =B\sum_{i=1}^{N}\sigma_{i}^{z},   
\end{gather}
where the subscript $i$ refers to the $i$th spin in the chain, and $\sigma_{i}^{k}$ denotes a Pauli spin operator with $k=x,y,z$. Without loss of generality, we have taken the magnetic field to point in the negative $z$ direction, such that the single-spin ground state is $\ket{\downarrow}$. We work in units where the magnetic moment, the lattice spacing, and $\hbar$ are all set to 1.

The second term in the Hamiltonian, $H_{g}$, defines pairwise interactions between different spins:
\begin{gather}\label{eq:Hg}
    H_{g} = -\sum_{i<j}g_{ij}\left[\sigma_{i}^{z}\otimes\sigma_{j}^{z}+\alpha \left(\sigma_{i}^{x} \otimes \sigma_{j}^{x} + \sigma_{i}^{y} \otimes \sigma_{j}^{y}\right)\right].
\end{gather}
Here, the interaction strength between spins $i$ and $j$ is given by $g_{ij}$, and we have encoded an anisotropy in the parameter $\alpha$, where $|\alpha|\leq 1$. For general $\alpha$, this model corresponds to the XXZ spin chain, while the particular values of $\alpha=0$ and $\alpha=1$ correspond to the Ising and XXX spin chains, respectively.

Whilst nearest-neighbour coupling is often assumed, next-nearest-neighbour coupling can better model some experimental compounds~\cite{Matsubara_1991}, and long-range interactions that decay as a power law can be manufactured~\cite{Smith_2016,Hung2016}. Thus, in general, we will consider either nearest neighbor (NN) or long-range (LR) interactions, respectively given by
\begin{subequations} \label{eq:g}
\begin{align}
    g^{\rm NN}_{ij} & = g \, \delta_{i, j-1} \, ,\\
    \label{eq:LR}
    g^{\rm LR}_{ij} & = \frac{g}{|i-j|^p} \, , 
\end{align}
\end{subequations}
where $g$ is a real constant and $p$ is a non-negative number. Note that the infinite-range case $g_{ij} = g$ corresponds to taking $p=0$ in Eq.~\eqref{eq:LR}, which is similar in spirit to the scenario considered in Ref.~\cite{Ferraro2017}. We assume attractive interactions ($g\geq0$), such that the ground state of the static Hamiltonian is ferromagnetic, \emph{i.e.}, $\rho_\downarrow  = \ket{\downarrow} \bra{\downarrow}^{\otimes N}$, and we always take this to be our initial state. In this case, the initial energy corresponds to 
\begin{align}
\tr\left[H_{0}\rho_\downarrow\right] & =\tr\left[H_{B}\rho_\downarrow\right]+\tr\left[H_{g}\rho_\downarrow\right]\nonumber\\
 & =-NB-\sum_{i<j}g_{ij}.
 \label{eq:initE}
\end{align}

In order to impart energy to the system, we consider a charging Hamiltonian $V$. In contrast to previous works, where $N$ batteries are collectively charged by an interacting potential, here a single many-body battery (with internal interactions) is charged using a  \emph{local} external driving field that changes the energy splitting Hamiltonian from $H_B$ to
\begin{gather}\label{eq:potential}
V =\omega\sum_{i=1}^{N}\sigma_{i}^{x},
\end{gather}
which is perpendicular to the original Zeeman splitting and uniform in space. This could be physically generated by imposing another external magnetic field, or by simply rotating the system relative to the existing field. By considering a local-only charging, we can isolate any entanglement generation during the charging as arising from the interactions within the spin chain itself.

The charging potential can, in general, be time dependent; however, we will consider the simpler scenario where it is constant during the charging interval. For $0<t<T$ we thus have the total Hamiltonian
\begin{gather}
H =H_{g}+V,\label{eq:totalH}
\end{gather}
which generates the unitary evolution
\begin{gather}
U_t  = \exp \left[-i \left(H_{g}+V \right) t \right].
\end{gather}
Hence if $\rho_\downarrow$ is the battery's initial state, then its state at time $t$ is $\rho_{t} =U_t \rho_\downarrow U^{\dagger}_t$. The deposited work is the difference in internal energy
\begin{gather}
W(t) =\tr\left[H_{0} \rho_{t} \right]-\tr\left[H_{0} \rho_\downarrow \right],\label{eq:work}
\end{gather}
and we have the total work $W\equiv W(T)$ at the end of the charging, such that the average charging power is simply
\begin{gather}
P =\frac{W}{T}.
\end{gather}

For $N$ independent spins (equivalently $g_{ij}=0$), the maximum total work scales as $W_{\rm ind}=N W^{\left(1\right)}$ (`ind' stands for `independent'), where $W^{\left(1\right)}$ is the work deposited on one spin; and similarly, the average power scales as $P_{\rm ind} = N P^{\left(1\right)}$. Below, we analytically and numerically compute the charging power for different parameters in the interaction Hamiltonian $H_g$.

\section{Local charging of a many-body battery}
\label{sec:local_charging_results}

For a single spin, the driving field in Eq.~\eqref{eq:potential} produces the following work at time $T$:
\begin{align} \label{eq:W1}
W^{\left(1\right)}=  
2B 
\sin^2 \left(\omega T\right).
\end{align}
The maximum work is therefore $W^{\left(1\right)} = 2B$, which corresponds to charging the spin from $\ket{\downarrow}$ to $\ket{\uparrow}$. The maximum power is $P^{\left(1\right)} \simeq 1.4 \, \omega B$ when $T\simeq 1.2/\omega$. Note that the time of maximum work and that of maximum power deposition do not coincide.

We now proceed to consider the charging of a many-body battery, where the constituent spins interact pairwise. In general, the unitary evolution according to the Hamiltonian of Eq.~\eqref{eq:totalH} cannot be solved analytically. However, as we describe in this section, we can gain insight into the problem by studying limiting cases where we can obtain analytic results. Therefore, in the following we analyze both the case of weak and strong interactions compared with the charging field strength. We begin this section by investigating the role of symmetries in the Hamiltonian, as this has important implications for whether interactions can affect the charging at all.

\subsection{The role of symmetries}

The most symmetric scenario we can consider is that of a spin chain with isotropic couplings, \emph{i.e.}, $\alpha=1$. Despite the quantum nature of its constituent components, this scenario generates no quantum correlations or entanglement, nor any effect of the interactions between the spins. To see this, first note that $V$ commutes with the interaction part of the Hamiltonian, $H_g$. Hence, the unitary evolution decomposes via the Baker–-Campbell–-Hausdorff formula to $U_t = e^{-iVt} e^{-iH_gt}$, leading to the final state
\begin{gather}
\begin{split}
    \rho_T &= e^{-iVT} \rho_\downarrow  \ e^{iVT} \\
    &= \bigotimes_{k=1}^N  e^{-i\omega \sigma_k^x T} \ket{\downarrow}\bra{\downarrow}_k  e^{i\omega \sigma_k^x T},
\end{split}
\end{gather}
which has no dependence on the interactions. 
Finally, one can show that $\tr\left[H_{g} \rho_{t} \right] = \tr\left[H_{g} \rho_{\downarrow} \right]$, which implies that the deposited work in Eq.~\eqref{eq:work} is also independent of the interactions.

\begin{figure}[t]
\begin{centering}
\includegraphics[width=1\columnwidth]{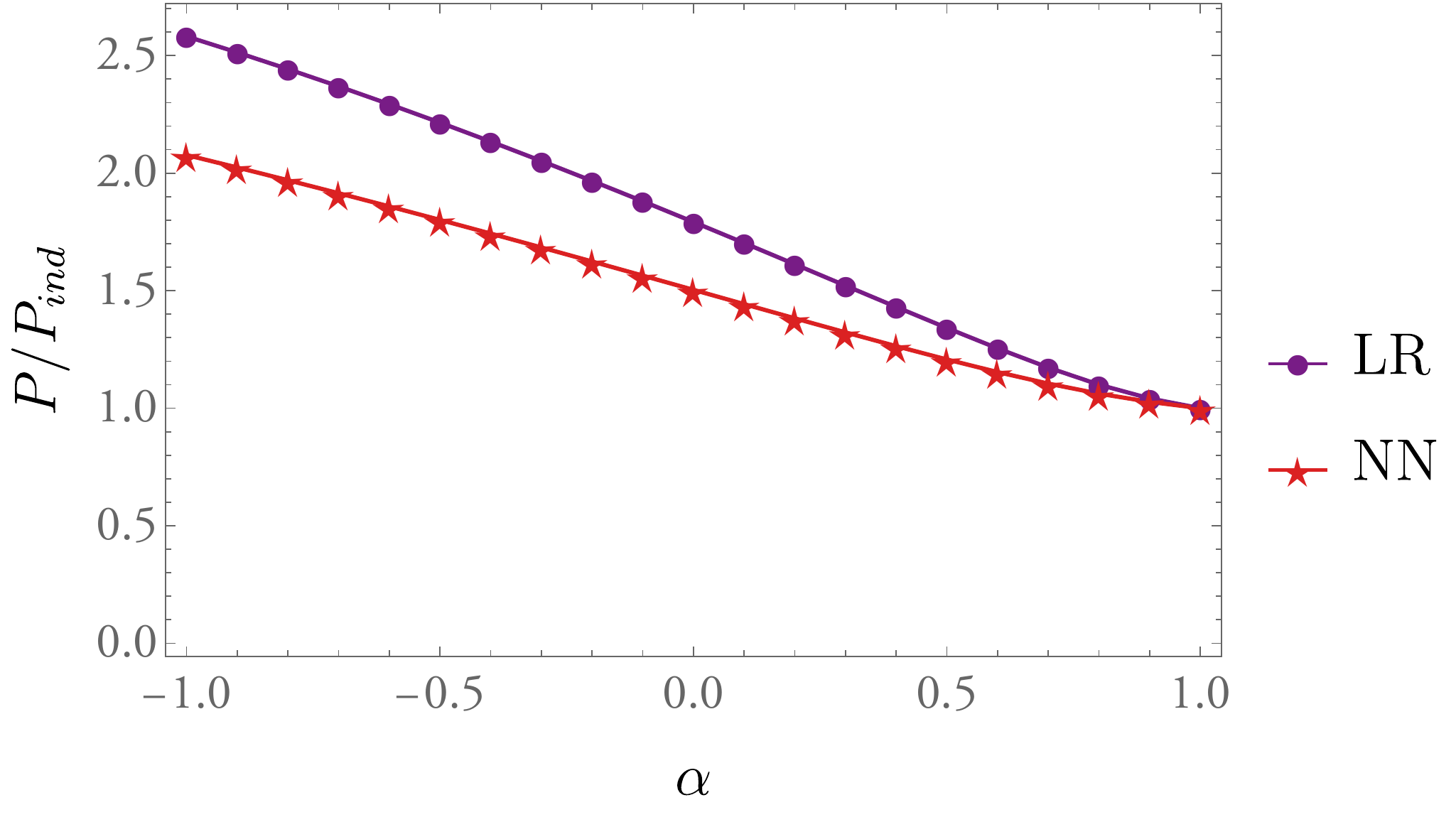}
\par\end{centering}
\caption{The maximum power (maximized over time) achievable by a spin chain of length $N=4$ as a function of anisotropy parameter $\alpha$ [see Eq.~\eqref{eq:Hg}]. For this illustration, we consider long-range (LR) interactions with $p=1$, as well as nearest neighbor (NN) interactions [see Eq.~\eqref{eq:g}]. The remaining parameters are chosen as $g=B$ and $\omega=4B$. $P_{\rm ind}$ is the power achieved with isotropic interactions, $\alpha=1$, corresponding to that of 4 independent spins. \label{fig:anisotropy}}
\end{figure}

This is our first result: that a spin chain with isotropic coupling leads to completely independent charging of each spin. In other words, the XXX Heisenberg spin chain will charge as though it were a collection of $N$ independent spins, regardless of the range or strength of the coupling interactions $g_{ij}$, and despite the fact that the added interactions change the spectrum of both the static and the charging Hamiltonians. In fact, we can see numerically that reducing the symmetry (\emph{i.e.}, changing $\alpha$ from 1) leads to a direct increase in maximum average power. As Fig.~\ref{fig:anisotropy} illustrates, the maximum power obtainable in the charging increases as the anisotropy increases. Correspondingly, as a direct consequence of the many-body nature of the interacting spin battery, the anistropic XXZ spin chain achieves a much higher power than the isotropic XXX spin chain.

Therefore, to take advantage of the spin-chain battery's capability of intrinsic many-body interactions, we must break a rotational or translational symmetry. This could also be achieved by applying a different charging field to each spin, and, to this end, a physically reasonable charging scheme is one where charging fields are only applied to one or both ends of the spin chain. However, our preliminary results suggest that only the end spin(s) and those connected via a direct interaction will charge, while the remainder of the spin chain remains uncharged. Indeed, this scenario holds even when the symmetries in $H_0$ are broken. Thus, a quantum advantage in this scheme is limited to short chains, or chains with long-range interactions where $p$ in Eq.~\eqref{eq:LR} is sufficiently small. Hence, we only consider schemes that charge all the spins in the chain uniformly. We will therefore focus on systems with broken rotational symmetry, \emph{i.e.}, we consider anisotropic spin-spin interactions. 

\subsection{Emergence of $N$-body interactions in the strongly interacting spin chain}\label{sec:schriefferwolf}

We now turn to the limit of a strongly interacting spin chain, which---as we shall explicitly demonstrate---allows us to identify emergent $N$-body interactions, even though our Hamiltonian only features a local charging field and is limited to two-body interactions. Our starting point is the observation that in the limit of strong attractive interactions, \emph{i.e.}, $g\gg\omega$, the Hamiltonian features a large spectral gap between the low-energy states $\rho_\downarrow =\ket{\downarrow} \bra{\downarrow}^{\otimes N}$ and $\rho_\uparrow =\ket{\uparrow} \bra{\uparrow}^{\otimes N}$, and all other states. The size of this gap generally increases with the range of the interactions and it depends on the anisotropy parameter $\alpha$. Taking, for concreteness, $\alpha=0$, we see that even if we limit ourselves to nearest neighbor interactions quantified by the interaction constant $g$, the size of the gap $\sim 2g$. This large separation of scales allows us to adiabatically eliminate the high-energy degrees of freedom~\cite{arXiv:1503.01369}, and project onto an effective low-energy Hamiltonian which acts only in the space spanned by $\rho_\downarrow $ and $\rho_\uparrow$.

\begin{figure}[t]
\centering
\includegraphics[width=.9\columnwidth]{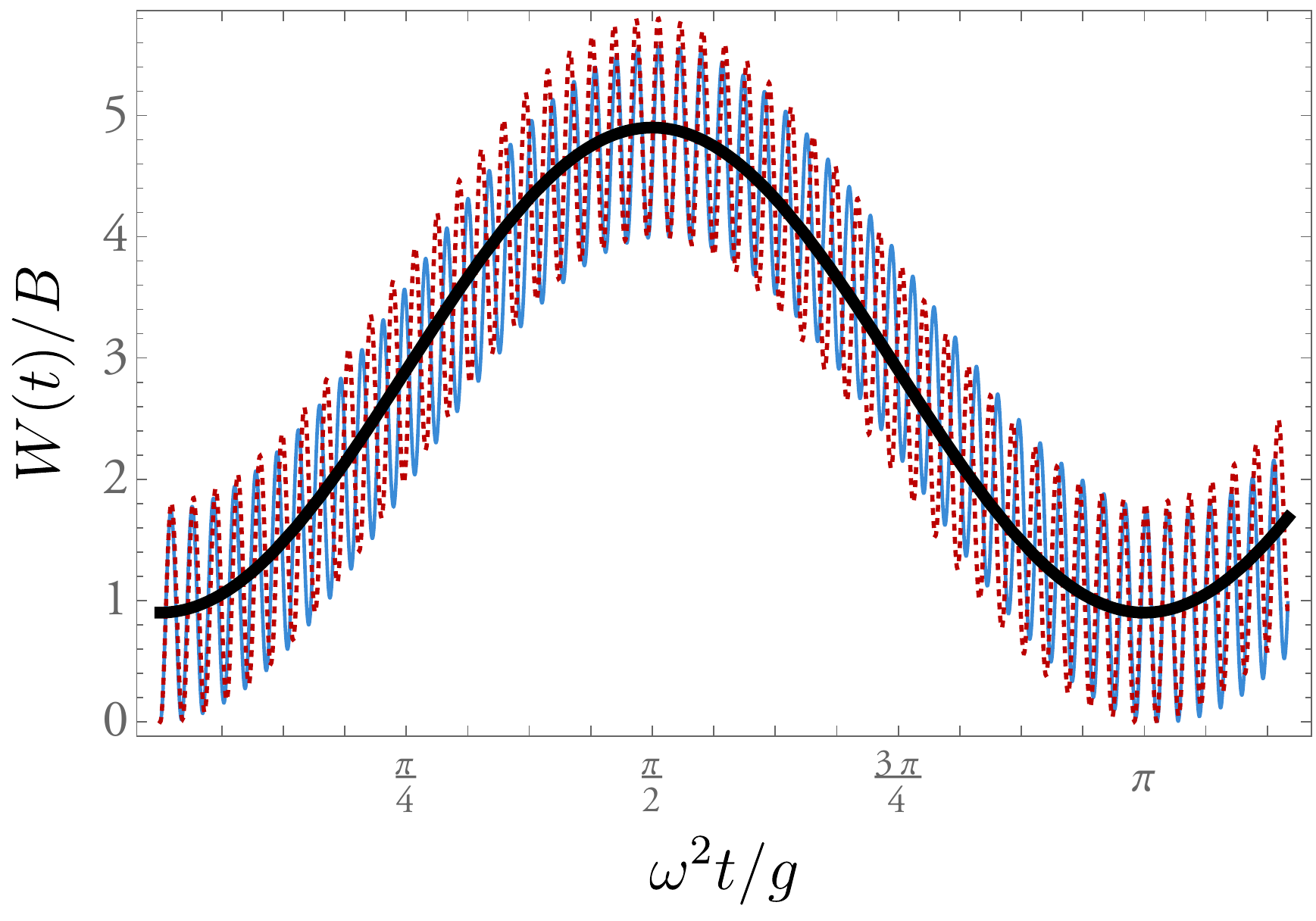}
\caption{Work as a function of time for a two-spin battery in the strong coupling regime (for this illustration we take, $\omega=3B$, $\alpha=0$, and $g=20B$). The exact result is shown as the solid blue curve. The slow oscillations, Eq.~\eqref{eq:slow}, related to the effective low-energy Hamiltonian~\eqref{eq:effH2}, are shown as a black solid line, while the sum of the slow and the fast oscillations of Eq.~\eqref{eq:fast} are shown as dotted red.}
\label{fig:2body}
\end{figure}

To see how this projection works, it is instructive to first consider the two-spin problem where, of course, there is no difference between short- and long-range interactions. In this particular case, we can take advantage of how the total spin $S$ commutes with both terms in the Hamiltonian, and thus remains constant during the charging. It is then straightforward to evaluate the time evolution analytically within the set of triplet states: $\ket{\uparrow\uparrow}$, $\frac1{\sqrt{2}}(\ket{\uparrow \downarrow}+\ket{\downarrow \uparrow})$, and $\ket{\downarrow\downarrow}$. While the work deposited onto the battery is in general a complicated function of time, it can essentially be decomposed into two regimes. The first is a regime dominated by the 
high-energy part of the spectrum, leading to fast oscillations:
\begin{gather}
    W_{\rm fast}(t)\simeq \frac{4\omega^2}{g}\sin^2(gt), \hspace{5mm} t\ll \frac{g}{\omega^2}.
\label{eq:fast}
\end{gather}
This result indicates that, by turning off the field at the first peak of these oscillations, we can charge the battery up to a maximum work $\sim \omega^2/g$ and a corresponding maximum power $\sim\omega^2$. At longer times, we find a second oscillatory behavior
\begin{gather}
    W_{\rm slow}(t)\simeq \frac{2\omega^2}{g} + 4 B \sin ^2\left(\omega^2t/g\right), \hspace{5mm} t\gg 1/g,
\label{eq:slow}
\end{gather}
where the first term arises from averaging out the fast oscillations. As seen by the prefactor $\sim B$, this slow oscillation of the work corresponds exactly to the charging between the states $\rho_\downarrow $ and $\rho_\uparrow$. Therefore, in this case the maximum work scales with $B$ while the maximum power goes as $B\omega^2/g$. We thus see that the achievable power is in general much greater if one uses the fast oscillations for charging, at the expense of depositing less work. The fast and slow oscillations are illustrated in Fig.~\ref{fig:2body}.

The slow oscillations found in Eq.~\eqref{eq:slow} may be understood as arising from an effective low-energy charging Hamiltonian for the two-spin problem
\begin{gather}
H_{\rm eff}=\frac{\omega^2}g\left[\ket{\uparrow \uparrow}\!\bra{\downarrow\downarrow}+\ket{\downarrow \downarrow}\!\bra{\uparrow\uparrow}\right],
\label{eq:effH2}
\end{gather}
which can be found using second order perturbation theory (we ignore a term proportional to the identity which does not lead to work deposition). This Hamiltonian acts solely within the space of $\rho_\downarrow $ and $\rho_\uparrow$, and as such the corresponding evolution proceeds through an entangled subspace.

\begin{figure}[t]
\begin{centering}
\includegraphics[width=1\columnwidth]{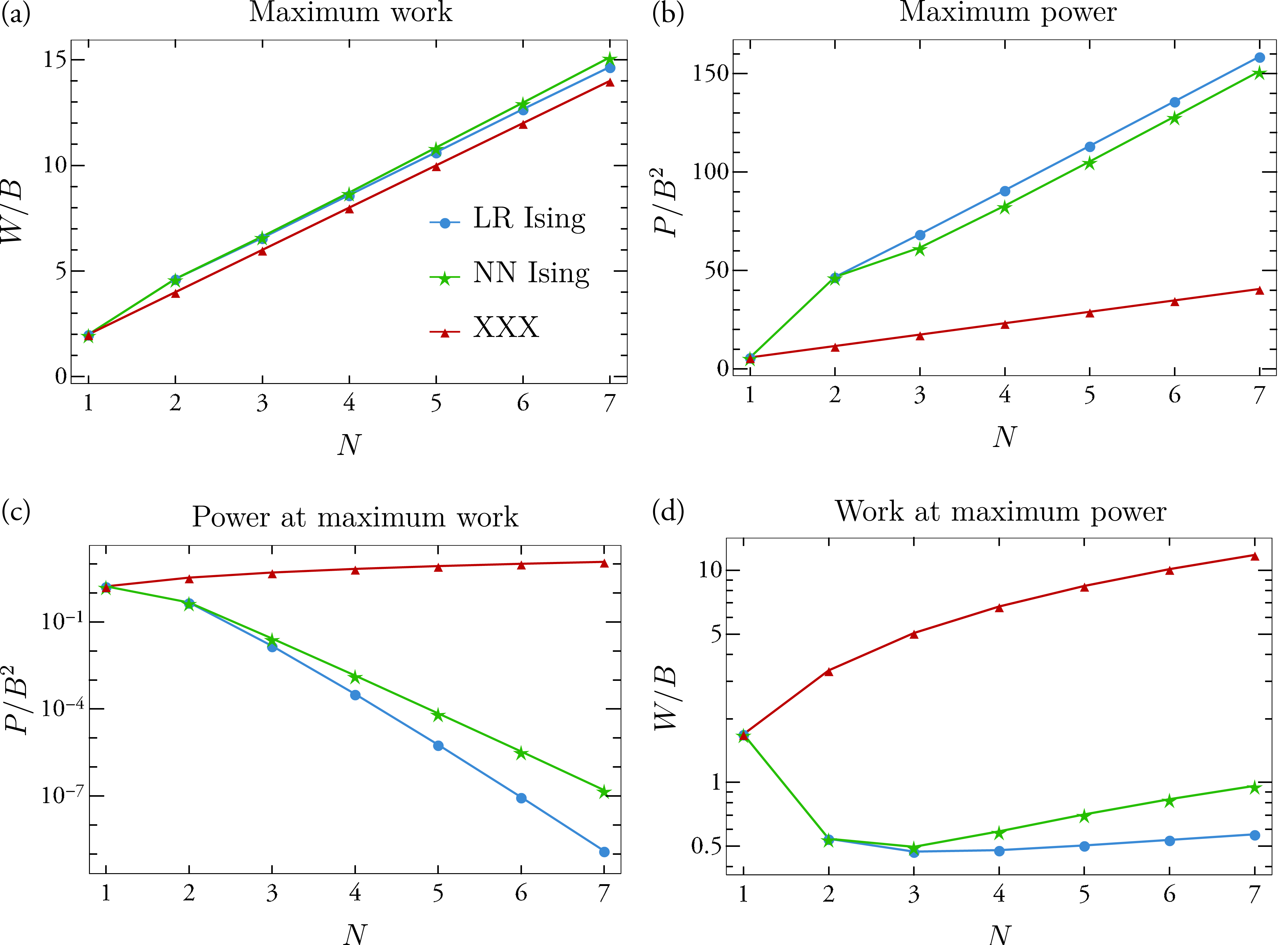}
\par\end{centering}
\caption{(a) Maximum achievable work and (b) power as a function of the spin-chain length. We illustrate this for isotropic (interaction independent) spin chains (XXX), and for anisotropic ($\alpha=0$) nearest neighbor (NN) interactions and long-range (LR) interactions with  $p=1$. For comparison, in (c) the average power at maximum work is shown, and in (d) the work achievable at maximum power. For this illustration, we take $g=100 B$ and $\omega=4B$.
\label{fig:numerics_strong}}
\end{figure}

We can generalize our results to longer spin chains in the regime of strong interactions. For general $N$, one also obtains an effective global entangling Hamiltonian similar to Eq.~\eqref{eq:effH2}, {\em i.e.}, of the form $\bigotimes_{k=1}^N  \ket{\downarrow}\bra{\downarrow}_k+h.c.$, and we find that the prefactor scales as $\omega^N/g^{N-1}$ with a coefficient that depends on the range of the interactions. This scaling emerges naturally within a perturbative approach in the small parameter $\omega/g$ by enumerating the virtual processes needed to connect the states $\rho_\downarrow$ and $\rho_\uparrow$ for general $N$. For instance, if one considers a chain consisting of three spins, each interacting with coupling constant $g$, then we find $W_{\rm slow}(t)=6B\sin^2(3\omega^3t/8g^2)$ arising from the effective Hamiltonian $H_{\rm eff}=\frac{3\omega^3} {8g^2} \left[ \ket{ \uparrow \uparrow \uparrow }\! \bra{ \downarrow \downarrow \downarrow } + \ket{ \downarrow \downarrow \downarrow} \!\bra{\uparrow \uparrow \uparrow} \right]$. This scaling means that for $N$ spins the maximum power due to the slow oscillation scales as $B\omega (\omega/g)^N$. These scalings are exemplified in Fig.~\ref{fig:numerics_strong}, where we show the maximum achievable power and work (corresponding to the first peaks of fast and slow oscillations, respectively) as a function of $N$ for a particular choice of parameters. Our results, which are calculated by exactly solving the system numerically, are also shown both for nearest neighbor interactions and the case of long-range interactions with $p=1$. We do not find a strong dependence on the participation number.

We can conclude that, although strong coupling can lead to effective many-body interactions in our model, the magnitude of such effective interactions decreases with our ability to produce them. The power deposited in the battery is actually worse when the spins traverse the correlated shortcut suggested in Ref.~\cite{Binder_2015}, and it becomes vanishingly small in the limit of a large number of spins.  Therefore, we now consider whether weaker interactions can lead to a faster charging of the many-body battery.

\subsection{Weakly interacting spin chain}

In the regime where the interactions are small compared to the driving strength $\omega$, {\em i.e.}, $G\equiv\sum_{i<j}g_{ij}\ll N\omega$, we can treat the interactions as a perturbation and derive approximate analytical results for the work deposited and charging power.

In the following, we simplify the derivation by separating the Hamiltonian into $V$ and $H_{g}$, and moving into the interaction picture with respect to the former. This is convenient since $V$ consists of only local terms whose spectral decomposition is straightforward. The interaction-picture density operator is then $\tilde{\rho}_t =e^{iVt} \rho_t  e^{-iVt}$,
with the corresponding interaction-picture Hamiltonians $\tilde{H}_{x,t} =e^{iVt}H_{x} e^{-iVt}$ where $x \in \{0,B,g\}$.
To first order in $G$, the first term of the deposited work, Eq.~\eqref{eq:work}, can be decomposed into:
\begin{align}
\tr\left[H_{0}\rho_t\right]
 \! \simeq \! \tr\left[\tilde{H}_{0,t} \rho_\downarrow \right]\label{eq:finalenergy}  \!\! - \! i\tr \bigg[\tilde{H}_{B,t} \textstyle{\int_{0}^{t}}  ds\left[\tilde{H}_{g,s},\rho_\downarrow\right]\bigg],
\end{align}
where we discard any terms of order $G^{2}$ and higher. 

Since $V$ is a local Hamiltonian, we can write $e^{\pm iVt} = \bigotimes_{j=1}^{N} e^{\pm i \omega t \sigma_j^{x}}$. The first part of Eq.~\eqref{eq:finalenergy} then becomes 
\begin{align}
\begin{split}
\tr\left[\tilde{H}_{0,t}\rho_\downarrow\right]  \simeq& -B \sum_{i=1}^{N} \cos\left(2t\omega\right)\\
 &-G \left[\alpha \sin^2\left(2t\omega\right)+\cos^2\left(2t\omega\right)
 \right].
\end{split}
\end{align}
The second term,  $\tr\left[\tilde{H}_{B,t} \int_{0}^{t} ds \left[\tilde{H}_{g,s}, \rho_\downarrow \right]\right]$, turns out to be identically zero after a straightforward expansion of the commutator. Hence, the total work as a function of time is
\begin{gather}
W(t) \simeq 2BN\sin^{2}\left(\omega t\right)+(1-\alpha)G \sin^{2}\left(2\omega t\right).
\label{eq:WT}
\end{gather}
Note that we recover the work of $N$ non-interacting spins when the interactions are isotropic, \emph{i.e.}, if we have $\alpha=1$.

\begin{figure}[ht]
\centering
\includegraphics[width=.9\columnwidth]{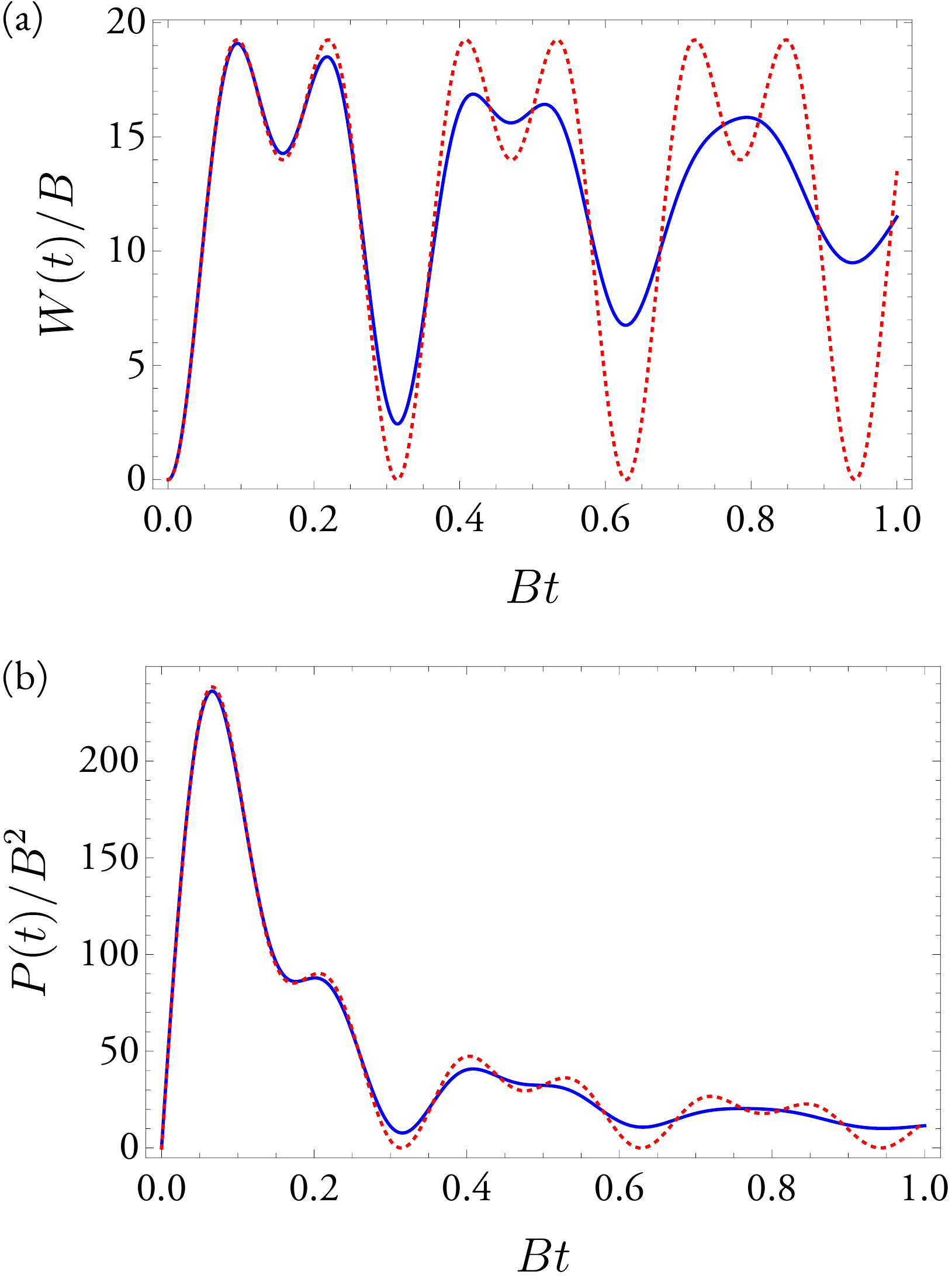}
\caption{(a) Work and (b) power as a function of time for a battery with 7 spins in the weak coupling regime (for this illustration we take $\omega=10B$, $g=B$, $\alpha=0$, and $p=1$). 
The perturbative result in Eq.~\eqref{eq:WT} is shown as a dotted (red) curve, while the exact result corresponds to the solid (blue) curve.}
\label{fig:weak}
\end{figure}

In Fig.~\ref{fig:weak}, we see that Eq.~\eqref{eq:WT} is a good approximation for the weakly interacting spin chain at short times $Gt <1$, which is sufficient for determining the maximum power and work. Since Eq.~\eqref{eq:WT} neglects higher order terms, it does not include the dynamical role of the interactions. Specifically, the dynamical effect of the interactions only appears at second order, and thus the period of oscillations in Eq.~\eqref{eq:WT} does not depend on~$G$. Instead, the first order perturbative work incorporates the energy stored in the interactions due to effective independent charging. Nevertheless, the last term of Eq.~\eqref{eq:WT} yields the potential for greater work deposition in an interacting many-body battery than in $N$ independent batteries. This is due to the existence of many-body eigenstates higher in energy than $\rho_\uparrow$ once $G$ is sufficiently large compared to $BN$.

In the regime where $2(1-\alpha)G/BN<1$, the maximum work is the same as in the independent case, with $W_{max,1} =2BN=W_{\rm ind}$ at $T_{max,1}=n\pi/2\omega$ (for any integer $n$). However, when $2(1-\alpha)G/BN>1$, these become local minima, as shown in Fig.~\ref{fig:weak}(a), and the maxima occur at times $T_{max,2} = \arccos \left\{-2BN/(4G(1-\alpha))\right\}/2\omega$, with corresponding work values
\begin{gather}
W_{max,2} =B^2N^2\dfrac{\left(1+2G(1-\alpha)/BN\right)^{2}}{4G(1-\alpha)}.
\end{gather}
In this case, the average power at maximum work is higher than in the independent case. 

The maximum average power in the weak coupling regime can be approximated as
\small
\begin{gather} \notag
P_{\max} =\max_{T}\left\{ 4 \omega \left[ \frac{BN}{2}+ (1-\alpha)G\cos^2(\omega T)\right] \dfrac{\sin^{2}(\omega T)
}{\omega T}\right\}
\end{gather}
\normalsize
which shows that the maximum power of the interacting chain is larger than its non-interacting counterpart when $\alpha \leq 1$. If the spin-spin interaction is finite-ranged, such as nearest neighbor or next-nearest neighbour, then $G\sim N$ and the power is enhanced by a constant factor only. This corresponds to a fixed participation number, and interaction order of $k=2$, which agrees with the charging power scaling ${\cal O}(mk^2) = {\cal O}(\text{constant})$ of Ref.~\cite{Campaioli_2016}. For long-range interactions \eqref{eq:LR}, we find that if the decay of the spin-spin coupling strength is sufficiently fast, \emph{i.e.} if $p>1$, then these long-range interactions can only provide, at most, an extra prefactor in the limit $N\rightarrow\infty$, with  $G$ converging to a constant factor. If we engineered a stronger  pairwise interaction  with $p=1$, then $G\sim N\log N$ and the power is now super-extensively enhanced as ${\cal O}\left(\log N\right)$. It is only when we have uniform magnitude infinite range coupling that we recover the scaling of Ref.~\cite{Campaioli_2016}, since $G\sim N^2$ and the spin chain's charging power enhancement is ${\cal O}(m) = {\cal O}(N)$. Such long-range interactions can, for instance, be engineered and controlled using atoms trapped in a photonic crystal waveguide~\cite{Hung2016}, thus highlighting the practical relevance for the model considered here. 

While we have framed these results as perturbative, they correspond exactly to the work and power achieved by switching off the interactions during charging, regardless of the parameter regime. That is, they represent the non-dynamical contributions of the interaction energy. We are able to achieve a super-extensive scaling even in this case, where the spins charge independently. This motivates a further study of the role played by the interactions when they can affect the dynamics.

\section{The Role Of Correlations} 
\label{sec:classical_chains}

\begin{figure}
\begin{centering}
\includegraphics[width=1\columnwidth]{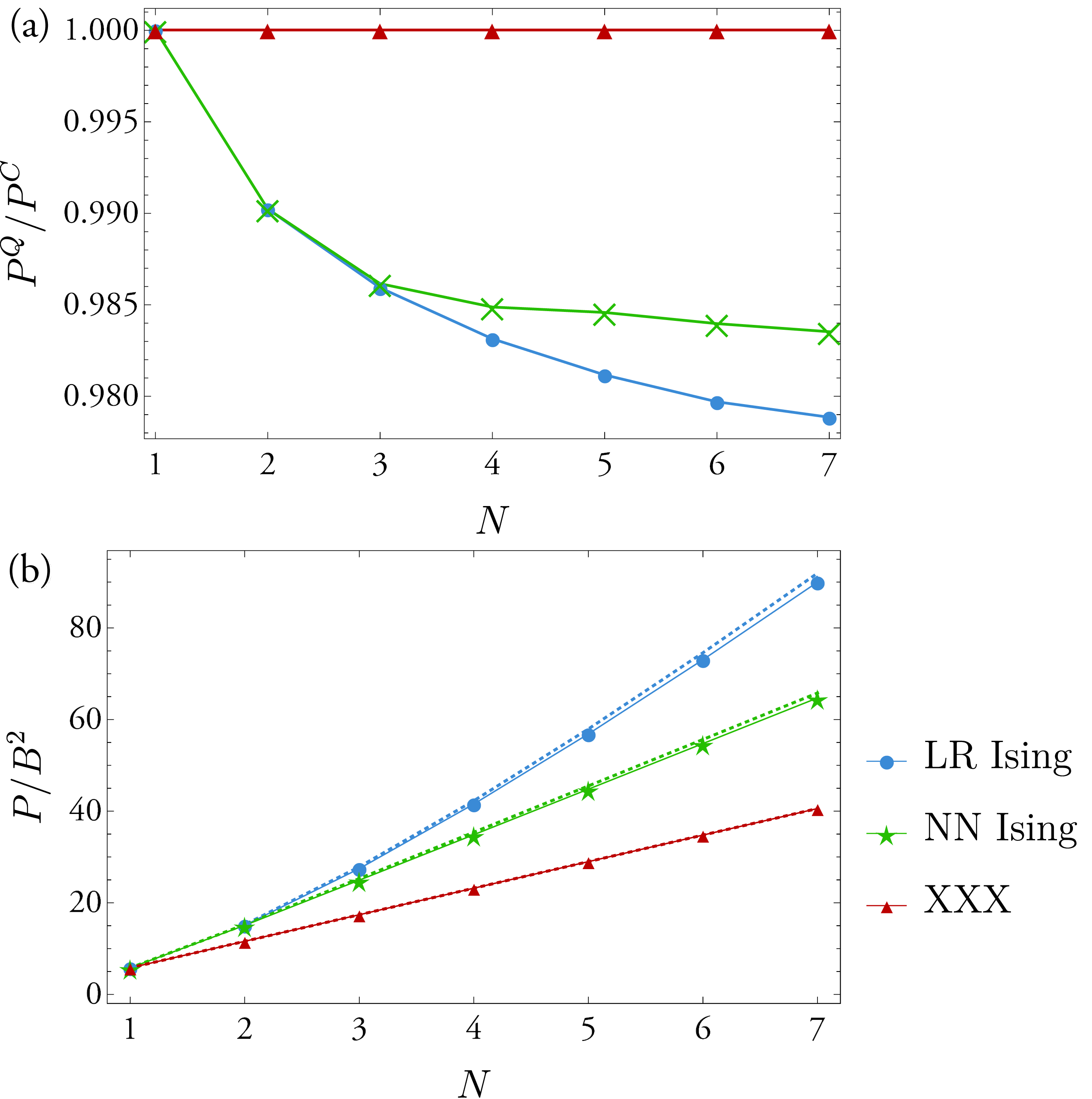}
\par\end{centering}
\caption{Comparison of the maximum power achievable for quantum (Q) and classical (C) spin chains of varying length. Here we consider isotropic (independent) spin chains (XXX) as well as maximally anistropic spin chains, $\alpha=0$, where we have long-range interactions with $p=1$ (LR Ising) and nearest neighbour (NN Ising) interactions. In (b), the solid line gives the quantum spin chain results, and the dotted line the classical spin chains. For this illustration, we have taken $g=B$ and $\omega=4B$.\label{fig:classicalspinchain}}
\end{figure}

Thus far, we have demonstrated that spin chains with anisotropic interactions can achieve greater power than in the non-interacting case. However,
a crucial question to ask is how much of this advantage derives from quantum correlations and entanglement? For example, in the  work by Binder \emph{et al.}~\cite{Binder_2015}, the quantum battery achieved greater power by taking a shortcut through an entangled subspace. To investigate this further, in this section, we compare the full dynamics of the spin chain with that of a mean-field model, where interactions cannot generate correlations.

More precisely, we consider the case where spin $m=1,\ldots,N$  evolves according to a local, time-dependent effective Hamiltonian $H_{m,t}^{\rm eff}$,
\begin{gather}
H_{m,t}^{\rm eff}=\tr_{!m}\left[H\left(\id_{m}\otimes\tr_{m}\left[\rho_t\right]\right)\right],\label{eq:6:Heff}
\end{gather}
where $\tr_{!m}$ is the trace over all spins except spin $m$ and $H$ is the original quantum Hamiltonian. As such, the total effective Hamiltonian is $H^{\rm eff}_t = \sum_m H_{m,t}^{\rm eff}$. Given that no  correlations (classical or quantum) can now build up, the state of the spin chain is  simply described by positions of each spin's Bloch vectors $\vec{S}_{m}=\left(S_{m}^{x}, S_{m}^{y}, S_{m}^{z}\right)$, where $\left| \vec{S}_{m} \right|=1$. 

The (nonlinear)  evolution of the spin chain under $H^{\rm eff}_t$ is equivalent to the evolution of the Bloch vectors according to the classical Hamiltonian $H_C$, where Pauli operators in $H$ are replaced by components of $\vec{S}_{m}$~\cite{RobertsTompson1988, Evertz1996}:
\begin{gather} \label{eq:classical}
\dfrac{d\vec{S}_{m}}{dt}=2\dfrac{\partial H_{C}}{\partial\vec{S}_{m}}\times\vec{S}_{m}.
\end{gather}
Here, each spin is subject to the time-dependent potential generated by the rest of the chain, but without any quantum back action. While this classical spin chain ultimately evolves in a very different state space, it has the same energetics as the quantum spin chain from which it was constructed. That is, the average energy of equivalent configurations is identical.

The power comparison between a quantum spin chain and its corresponding classical mean-field model is shown in Fig.~\ref{fig:classicalspinchain}. As expected, the XXX spin chains---where the spins evolve independently---achieve the same maximum power in both classical and quantum versions, as no correlations were created in the quantum spin chain in the first place. However, when quantum correlations are created, the corresponding classical spin actually charges faster than the quantum one for the parameters we considered. This demonstrates that the correlations between spins are \emph{not} the important factor in improving the charging power of a many-body battery. Rather, it is the additional interaction energy between the spins that provides the boost in power.

For the case of infinite-range interactions ($p=0$), we can formally show that we obtain the classical mean-field model in the limit $N \gg 1$.
Defining the average spin operator $\tilde{s}_k = \frac{1}{N}\sum_j \sigma^k_j$, we can rewrite the interaction Hamiltonian as
\begin{align}
    H_g^{p=0} = -\frac{g N^2}{2} \left[\tilde{s}_z^2 + \alpha \left( \tilde{s}_x^2 + \tilde{s}_y^2 \right) \right] ,
\end{align}
where we have dropped an unnecessary constant term. Similarly, we can write $H_B = BN \tilde{s}_z$ and $V = \omega N \tilde{s}_x$. It is easy to show that the average spin operators obey the commutation relations
\begin{subequations}
\begin{align}
    [\tilde{s}_j,\tilde{s}_k] & = \frac{2i}{N} \epsilon_{jkl} \tilde{s}_l \\
   {\rm and}\quad [\tilde{s}^2,\tilde{s}_k] & = 0,
\end{align}
\end{subequations}
where $\tilde{s}^2 = \tilde{s}^2_x+\tilde{s}^2_y + \tilde{s}^2_z$ and $\epsilon_{jkl}$ is the Levi Civita symbol. Therefore, we see that these operators all commute in the limit $N \to \infty$, and thus behave like the Bloch vectors for a classical spin. Furthermore, $\tilde{s}^2$ commutes with all the terms in the Hamiltonian, and thus the magnitude of the spin is always conserved. Since we start in the state $\rho_\downarrow$, we have magnitude $|\tilde s| = \sqrt{\tr\left[\tilde{s}^2 \rho_\downarrow \right]} = \sqrt{1+ 2/N}$, which tends to 1 for large $N$, once again mimicking the behavior of a classical spin. 
Finally, we can derive the equations of motion for the spin operators:
\begin{subequations}
\begin{align}
    \frac{d\tilde{s}_x}{dt} & = g N (1-\alpha) \left(\tilde{s}_y \tilde{s}_z + \tilde{s}_z \tilde{s}_y  \right), \\
    \frac{d\tilde{s}_y}{dt} & = -2\omega \tilde{s}_z - g N (1-\alpha) \left(\tilde{s}_x \tilde{s}_z + \tilde{s}_z \tilde{s}_x  \right), \\
    \frac{d\tilde{s}_z}{dt} & = 2\omega \tilde{s}_y.
\end{align}
\end{subequations}
These are exactly the classical equations obtained from Eq.~\eqref{eq:classical} if we assume that $\tilde{s}_k$ all commute with one another. Note that the strength of the interactions simply scales linearly with $N$ in this case.

Hence, we have shown that the infinite-ranged interacting spin chain in the large-$N$ limit behaves like a global classical spin, where spin-spin correlations are absent. Given that the infinite-range case generates the largest enhancement of the charging power, this further supports the conclusion that correlations are unnecessary for the efficient operation of our many-body battery.

\section{Conclusions}\label{sec:Conclusions}

We have examined the viability of quantum spin chains as a platform for a many-body battery. Our setup differs from those previously considered, where multiple batteries were charged collectively by inducing interactions with an external charging field. Instead, the spin chains forming our battery include intrinsic interactions between the constituent spins, while the charging is achieved by applying a local external field. Nevertheless, as in previous studies, our aim has been to look for speed-ups in charging power. In particular, we have investigated the role that the interactions play in producing such speed-ups. We have found that, in our model, enhancements in charging power can be mainly attributed to the extra energy available, rather than the ability to traverse correlated regions of the state space.

In order to couple to the higher energy many-body eigenstates of the interacting spin system, we require the translational or rotational symmetry to be broken, and thus we have focused on anisotropic spin-spin interactions.
In the \emph{strong} interaction limit, we have demonstrated that the spins traverse the entangled subspace responsible for the speed-up in Ref.~\cite{Binder_2015}, thanks to the emergence of an effective Hamiltonian with an interaction term that involves all spins simultaneously. However, we have found that the strength of this effective Hamiltonian is inversely proportional to the coupling strength and vanishes  exponentially with the number of spins. While there can be an increase in power relative to the independent charging case, this comes at the expense of the amount of work deposited in the battery, which becomes negligible in this limit.

We next examined the \emph{weak} coupling regime. To first order in the coupling strength, we found that the interactions in the chain play no dynamical role, such that the spins effectively charge independently. Nevertheless, the interactions contribute to the work done and can lead to an increase in charging power in some regimes.
The independent nature of charging in this regime, combined with the poor performance of the effective many-body Hamiltonian mentioned above, led us to further investigate the role of correlations in the final section of this paper.  There, by comparing with dynamics in which no correlations were allowed to build up, we showed that mean-field effects can account for any increase in charging power coming from the interactions in our model.
In other words, any speed-ups we see arise from the increased energy experienced by each spin due to interactions with the other spins in the chain.
This scenario fundamentally relies on the interactions being intrinsic to the many-body battery, rather than being imposed temporarily during the charging process.

Quantum technologies typically aim for exponential, or at least quadratic, advantage over their classical counterparts. However, in many applications even a constant advantage (one that does not scale with the size of the quantum system) is desirable. We have shown that physically constrained quantum batteries have the potential for faster charging over their classical (non-interacting)  counterparts. Specifically, our work illuminates how the structure of the interaction Hamiltonian may be designed to build fast charging quantum batteries. Our work opens up the potential feasibility of spin-chain quantum batteries, and is a step towards combining the concepts of quantum thermodynamics with the practicality of condensed matter systems. We leave the inclusion of noise in our problem as a future exercise.

\begin{acknowledgments}
This work was supported by a Samsung Global Research Outreach grant. JL is supported through the Australian Research Council Future Fellowship FT160100244.
JL and MMP acknowledge support from the ARC Centre of Excellence in Future Low-Energy Electronics Technologies (CE170100039).
\end{acknowledgments}

\bibliography{biblio}

\end{document}